\documentclass[prl,aps,twocolumn,groupedaddress,showpacs,floatfix,noshowpacs]{revtex4}
\usepackage{amsmath,amssymb,multirow,epsfig,bm,color}
\usepackage{hyperref}
\usepackage{url}

\usepackage[a4paper]{geometry}
\newcommand{\avg}[1]{\langle #1 \rangle}

\newcommand{\beq}{\begin{equation}}
\newcommand{\eeq}{\end{equation}}
\newcommand{\bea}{\begin{eqnarray}}
\newcommand{\eea}{\end{eqnarray}}

\newcommand{\eF}{\varepsilon^{}_{F}}

\newcommand{\ssec}[1]{\vspace{1mm}\noindent\emph{#1}.---}

\newcommand{\abs}[1]{\lvert{#1}\rvert}            

\newcommand{\nup}{n_{\uparrow}}

\newcommand{\ndown}{n_{\downarrow}}

\begin{document}

\title{Suppressed solitonic cascade in spin-imbalanced superfluid Fermi gas}
\author{Gabriel Wlaz\l{}owski$^{1,2}$, Kazuyuki Sekizawa$^{1,2}$, Maciej Marchwiany$^{3}$, Piotr Magierski$^{1,2}$}

\affiliation{$^1$Faculty of Physics, Warsaw University of Technology, Ulica Koszykowa 75, 00-662 Warsaw, Poland}
\affiliation{$^2$Department of Physics, University of Washington, Seattle, Washington 98195--1560, USA}
\affiliation{$^3$Interdisciplinary Centre for Mathematical and Computational Modelling (ICM), University of Warsaw, A. Pawi\'{n}skiego 5a, 02-106 Warsaw, Poland}

\email{gabriel.wlazlowski@pw.edu.pl}

\begin{abstract} 
Cold atoms experiments offer invaluable information on superfluid dynamics, including decay cascades of topological defects. While the cascade properties are well established for Bose systems,  our understanding of their behavior in Fermi counterparts is very limited, in particular
in spin-imbalanced systems, where superfluid (paired) and normal
(unpaired) particles naturally coexist giving rise to complex spatial structure of the atomic cloud.  Here we show, based on a newly developed microscopic approach, that the decay cascades of topological
defects are dramatically modified by the spin-polarization.  We demonstrate that decay cascades end up at
different stages: ``dark soliton'', ``vortex ring'' or ``vortex line'', depending on the polarization.  We reveal that it is caused by sucking of unpaired particles into the soliton's internal structure.
As a consequence vortex reconnections are hindered and we
anticipate that quantum turbulence phenomenon can be significantly affected, indicating
new physics induced by polarization effects. 
\end{abstract}

\maketitle

At very low temperatures, both bosonic and fermionic gases become superfluid.
The superfluidity is directly related to Bose-Einstein condensation, an intrinsic property of bosonic systems, where all particles occupy the lowest energy state. They loose their individuality and start to behave collectively like one giant particle~\cite{BECPethick}. Thus, the description of bosonic superfluidity is as simple as the description of a single quantum particle. 
Situation changes dramatically in the case of fermionic systems. Due to Pauli principle, fermions cannot condensate in a single quantum state. However 
nature found workaround. Two fermions with opposite spins create a correlated state, known as a Cooper pair. The total spin of the pair is an integer, like for bosons, and thus they can create a condensate. The most standard scenario ($s$-wave superfluidity) is realized when all Cooper pairs 
condense in a zero momentum state. However,  
the size of a Cooper pair cannot be neglected in the description of fermionic superfluids,
which introduces serious complication in theoretical treatment. 
Moreover, fermionic superfluids exhibit various phenomena that are absent in bosonic systems.
For example, introducing spin-imbalance into a Fermi system triggers qualitatively new behavior, since
in this case the Cooper-pairing mechanism is frustrated.
This opens various possibilities for Cooper pairing and many scenarios including appearance of novel phases have been hotly debated over years~\cite{Radzihovsky,Chevy}.
The most typical, and experimentally observed, is a spontaneous spatial separation 
of a fully-paired superfluid component from unpaired particles being in a normal state~\cite{Zwierlein1,Zwierlein2,Partridge,Shin}. Here we show, for the first time, how the spin imbalance influences dynamics of various topological excitations in the superfluid component.

We study dynamics of spin-imbalanced fermionic systems by means of numerical experiments. We apply the currently most complete description of the problem, which can be offered by quantum theory of many-body systems.
This approach utilizes an extension of the density functional theory to superfluid systems~\cite{Oliveira,Wacker}. 
Further developments by Bulgac et al. (see e.g. \cite{ARNPS__2013}) resulted in an accurate and numerically tractable formulation of the theory achieved by introducing a local pairing field $\Delta(\bm{r})$. It gave rise to the approach known as time-dependent superfluid local density approximation (TDSLDA).
Application of this very accurate microscopic method
has its price: the method requires enormous computing resources which can only be provided by top-tier computers. In recent years, supercomputers achieved petascale performance and it allowed TDSLDA to be successfully applied to variety of realistic systems like ultra-cold atomic gases~\cite{Science__2011,PRL__2014,PRAR__2015}, atomic nuclei~\cite{PRL__CoulEx,PRL__2016,PRL_MSW_2017} or interiors of neutron stars~\cite{VortexPinning}. 
In this letter, we apply an extension of TDSLDA, 
known as time-dependent asymmetric superfluid local density approximation (TDASLDA), for spin-imbalanced unitary Fermi gas~\cite{LNP__2012,ASLDA-LOFF}. In our studies we simulate scenarios that can be created in laboratories by means of ultra-cold atomic gases of ${}^{6}$Li or ${}^{40}$K at Feshbach resonance. Precisely, we work with a unitary Fermi gas at very low temperature $k_B T\approx 0.01\,\eF$, which is well below the transition temperature ($k_B T_c\approx 0.15\,\eF$ for spin symmetric system), where $\eF=\hbar^2(6\pi^2n_{\uparrow})^{2/3}/2m$ is the Fermi energy of majority spin component of density $n_{\uparrow}$. The gas is confined in elongated 3D harmonic trap. For detailed information about simulation parameters and numerical implementation of the TDASLDA equations, see Supplemental Material~\cite{Supplemental}. 

\begin{figure*}[th]
\includegraphics[width=\textwidth, trim=0 0 0 0, clip]{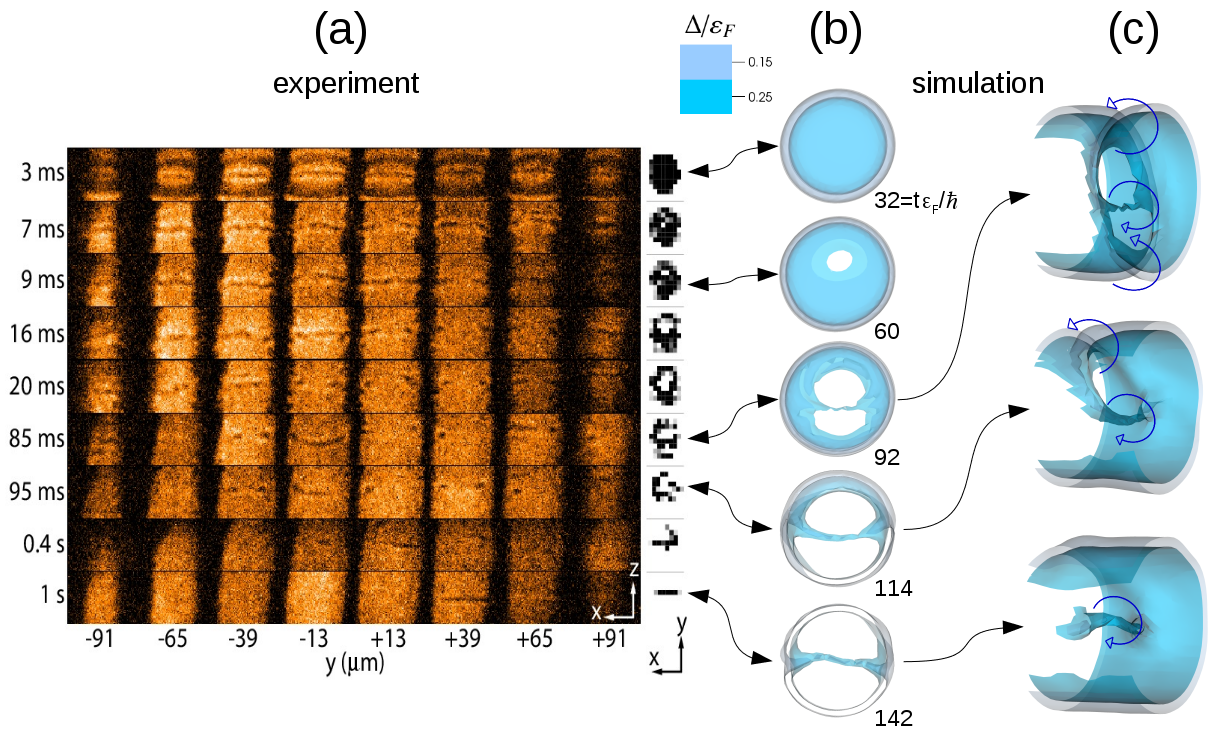}
\caption{(a) Time evolution of the ultra-cold atomic cloud after the phase imprint measured by the MIT experiment~\cite{SolitonicCascade}.  In right column, reconstructed spatial structure of topological defect  from the tomographic images is displayed. The figure has been reprinted 
with permission 
from: M.J.H. Ku et al., Phys. Rev. Lett. \textbf{116}, 045304 (2016).
(b) Selected frames from numerical simulation by TDASLDA showing consecutive stages of the solitonic cascade. Blue surfaces indicate topology of the order parameter $\Delta(\bm{r})$. By arrows we demonstrate that the numerical simulation provides the same time evolution of the system as seen in the experiment. For full movie see Supplementary Information. 
(c) Internal structures of the stages containing quantized vortices. The three frames contain, respectively, $\Phi$-soliton, deformed vortex ring and vortex line. Arrows indicate the gas circulation in vicinity of the topological defects.
\label{fig:1}}
\end{figure*}
Before we start to explore dynamics in spin-imbalanced systems, let us demonstrate accuracy of the method in the case of an unpolarized system. 
Last year, an experimental group at MIT~\cite{SolitonicCascade} provided very detailed information about time evolution of resonantly interacting ${}^{6}$Li atoms. The evolution was triggered by exposing one half of the harmonically-confined cloud of atoms to a blue-detuned laser beam resulting in changing of the phase of the complex-valued order parameter (associated with the superfluid phase) by $\pi$. Subsequent tomographic imaging allowed for
tracing the spatial structure of induced topological defects as a function of time. In Fig.~\ref{fig:1} we present comparison between experimental findings of Ref.~\cite{SolitonicCascade} and numerical results of TDASLDA with about $600$ atoms.
Although ratio of chemical potential to the trap angular frequency (in the transverse direction) $\mu/\hbar \omega_{\perp} \approx 3.5$ is by about order of magnitude smaller than in the experiment, we find a remarkable agreement between theory and the data. 
Both experiment and numerical simulation reveal the same stages of evolution with the same configurations of topological excitations present. 
Note that we have not adjusted any theory parameter to the experimental results. 
Interestingly, our simulation correctly reproduces a peculiar stage that visually resembles a ``$\Phi$-soliton", whose existence was predicted in \cite{Chladni_solitons} for traps where $\mu/\hbar \omega_{\perp} \gtrsim 2.9$. 
We confirmed that the flow pattern of the transient object is indeed consistent with expected pattern for a $\Phi$-soliton.
The results prove that dynamics of topological excitations within TDASLDA are correct, contrary to simplified models
based on Gross-Pitaevskii type equation~\cite{PRAR__2015}.

\begin{figure*}[th]
\includegraphics[width=\textwidth, trim=0 0 0 0, clip]{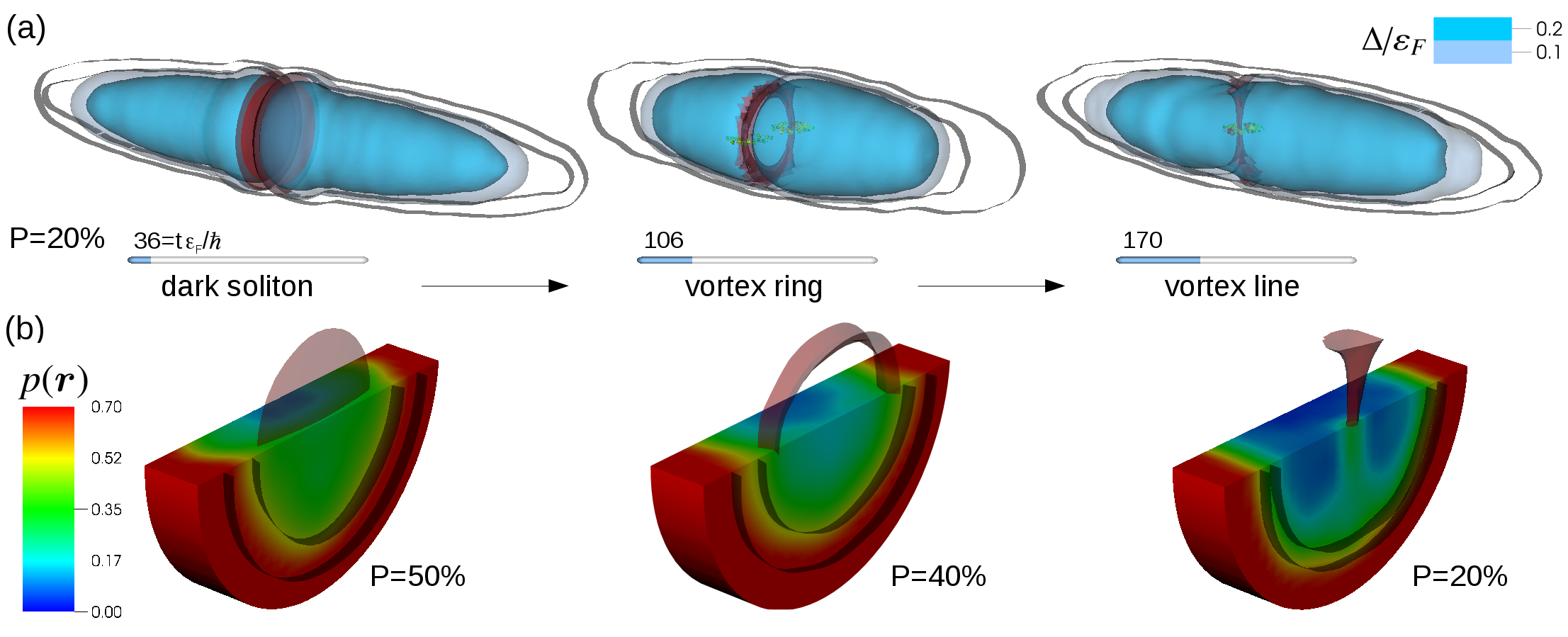}
\caption{(a) Selected stages of the solitonic cascade in spin-imbalanced cloud of atoms. Number of atoms is $N_{\uparrow}=304$ and $N_{\downarrow}=202$, which corresponds to the total polarization $P=20\%$. Blue contours show profile of the order parameter and define the volume where the condensate of Copper pairs resides. Regions containing topological defects are highlighted by red color. Two contours around the condensate indicate place where density of spin-down (internal contour) and spin-up (external contour) decreases by $90\%$ with respect to the value at the center of the cloud. Between them highly polarized gas in normal state exists. Full movie is attached to the Supplementary Information. (b) Final states of the cascade as a function of the cloud polarization $P$. Red contour indicates the topology of nodal plane of the order parameter. Slice of the cloud presents local polarization of the medium $p(\bm{r})=\frac{n_{\uparrow}(\bm{r})-n_{\downarrow}(\bm{r})}{n_{\uparrow}(\bm{r})+n_{\downarrow}(\bm{r})}$ in the vicinity of the topological defect. Two black contours indicate shell of the normal component, as defined above. Its thickness becomes wider as we increase the spin-imbalance of the cloud.
\label{fig:2}}
\end{figure*}
Having at hand the tool that is capable of describing correctly a solitonic cascade, we focus now on dynamics in spin-imbalanced systems.
We repeat simulations for the same setup, but now we vary the total polarization of the system $P=\frac{N_{\uparrow}-N_{\downarrow}}{N_{\uparrow}+N_{\downarrow}}$.
Namely, we fixed the number of spin-up atoms $N_{\uparrow}=304$ and the number of spin-down atoms was gradually 
decreased to get polarizations: $P=20\%,\, 40\%,\, 50\%$ and $60\%$.
As expected, low-energy initial states reveal configurations with 
phase separation. The superfluid component with marginal local polarization $p(\bm{r})=\frac{n_{\uparrow}(\bm{r})-n_{\downarrow}(\bm{r})}{n_{\uparrow}(\bm{r})+n_{\downarrow}(\bm{r})}\approx 0$
is located inside the cloud, whereas a highly polarized ($p(\bm{r})\approx 1$) cloud of unpaired particles 
is expelled outside, forming a shell of normal state, where $n_{\sigma}(\bm{r})$ denotes the local density of particles with spin $\sigma=\{\uparrow,\downarrow\}$, see Supplemental Material~\cite{Supplemental} for figures. 
In each case, after applying the phase imprint procedure, a dark soliton is created.
The most striking observation is that the polarization affects stability of the topological defects. 
In Fig.~\ref{fig:2}(a) we present consecutive stages of the solitonic cascade for the case of total polarization $P=20\%$, while Fig.~\ref{fig:2}(b) demonstrates the final product of the cascade for different polarizations. 
For relatively low polarization ($P=20\%$), we observe a full solitonic cascade that ends up with the vortex line. Once we increase the polarization up to $P=40\%$, we observe that the vortex ring is stabilized. It moves towards one edge of the cloud, where it collapses  
and re-forms a dark-soliton. The soliton exhibits the same instability and seeds a new vortex ring of the opposite circulation but with larger radius which propagates towards another edge of the cloud. Since the size of the ring is comparable with the size of the condensate, the superflow is largely affected by the presence of the shell of normal particles. As a consequence the vortex ring rapidly dissipates its energy to the normal component
and is removed from the cloud. Finally, for polarization $P=50\%$ and $60\%$ we observe formation of the dark soliton only. The soliton propagates towards the edge of the cloud and vanishes, indicating suppression of the snake instability. Phenomenon of stabilization of a dark soliton due to spin-imbalance was studied in papers~\cite{ReichlMueller,Lombardi2} under hypothesis of system uniformity---trap effects were completely neglected and dynamics were investigated effectively in 2D. 
Here we confirm, that the effect survives in full 3D calculations, which is not a~priori obvious, as dimensionality as well as trap effects play a very important role in the context of stability of topological defects. A stable soliton, filled with excess of fermions, can also be regarded as one limit of the long-sought Fulde-Ferrell-Larkin-Ovchinnikov state~\cite{SolitonicCascade,VarennaZwierlein,Dutta}.

One may be tempted to conclude  
that the observed stabilization effects are solely due to the geometry of the condensate: the size of the condensate decreases as the polarization increases. As is well known, stability of  
topological defects depends on the geometry of the condensate, e.g., dark solitons are stable in 1D, indicating that it is stabilized when the size of the cloud is sufficiently small. To clarify the main cause of the stabilization effects, we performed an additional simulation for a spin-balanced system with $N_{\uparrow}=N_{\downarrow}=76$. For the latter case, the size of the condensate is comparable to the case where $N_{\uparrow}=304$ and $N_{\downarrow}=76$, i.e., the $P=60\%$ case. We found that the dark soliton in the spin-balanced system is not stable and decays to a quantized vortex. Based on this observation, we conclude that the polarization effects are responsible for suppression of the solitonic cascade.

Let us now investigate internal structure of 
created topological defects. In Fig.~\ref{fig:2}(b) we present the local polarization in the vicinity of various topological defects. Clearly, the defects exhibit finite polarization at level $p(\bm{r})\approx 25$--$35\%$, irrespective of the total polarization. It means that the defects contain more spin-up particles than the surrounding medium. 
The effect has also been observed within simplified models~\cite{Lombardi,HuLiuDrummond}. The change of internal structure influences the rigidity of the defects
and consequently affects its dynamics. This can be understood based on energetic considerations. Energy contained in a defect is related to its volume $V$, which can be estimated as $\pi R^2\cdot\xi$, $2\pi R\cdot\pi\xi^2$ and $2R\cdot\pi\xi^2$ for the dark soliton, the vortex ring and the vortex line, respectively, where $R$ is axial radius of the cloud and $\xi$ is the coherence length. Once they decay solitonic defects decrease their internal energy, 
partially due to the decrease of the volume (there is no direct proportionality between the defect energy and its volume, as these objects arise from non-linearities in the TDASLDA
equations). In polarized medium, the defect carries excess of unpaired particles that can be treated as free Fermi gas. When the defect changes its topology, the excess particles have to be stored in smaller volume which increases its density and related energy. 
Thus, we have the interplay of the two effects and the change of the topology is not always energetically favorable. There is also another possibility, namely, that excess of spin-up particles diffuses from the interior of the topological defect to the outer shell in order to allow for modification 
of its topology. However this process is expected to be very slow, as the gas is characterized by a very low spin diffusion coefficient~\cite{Sommer,WlazlowskiQMC}. 
Similar phenomenon, where a dark soliton is stabilized due to change of its internal structure, was observed experimentally in two-component Bose-Einstein condensates (BECs)~\cite{Andersonetal}. Also, configurations where  one  component  of  such  system  supports  a  vortex state and the other is filling the vortex core have been observed~\cite{Matthewsetal,Andersonetal2}.  In general two-component BECs exhibit a rich variety of configurations whose stability depend on mutual interactions between the superfluid components~\cite{Kasamatsu}, while in our case the stability is related to interplay between condensed Cooper pairs and unpaired fermions.

\begin{figure}[t]
\includegraphics[width=\columnwidth, trim=0 0 0 0, clip]{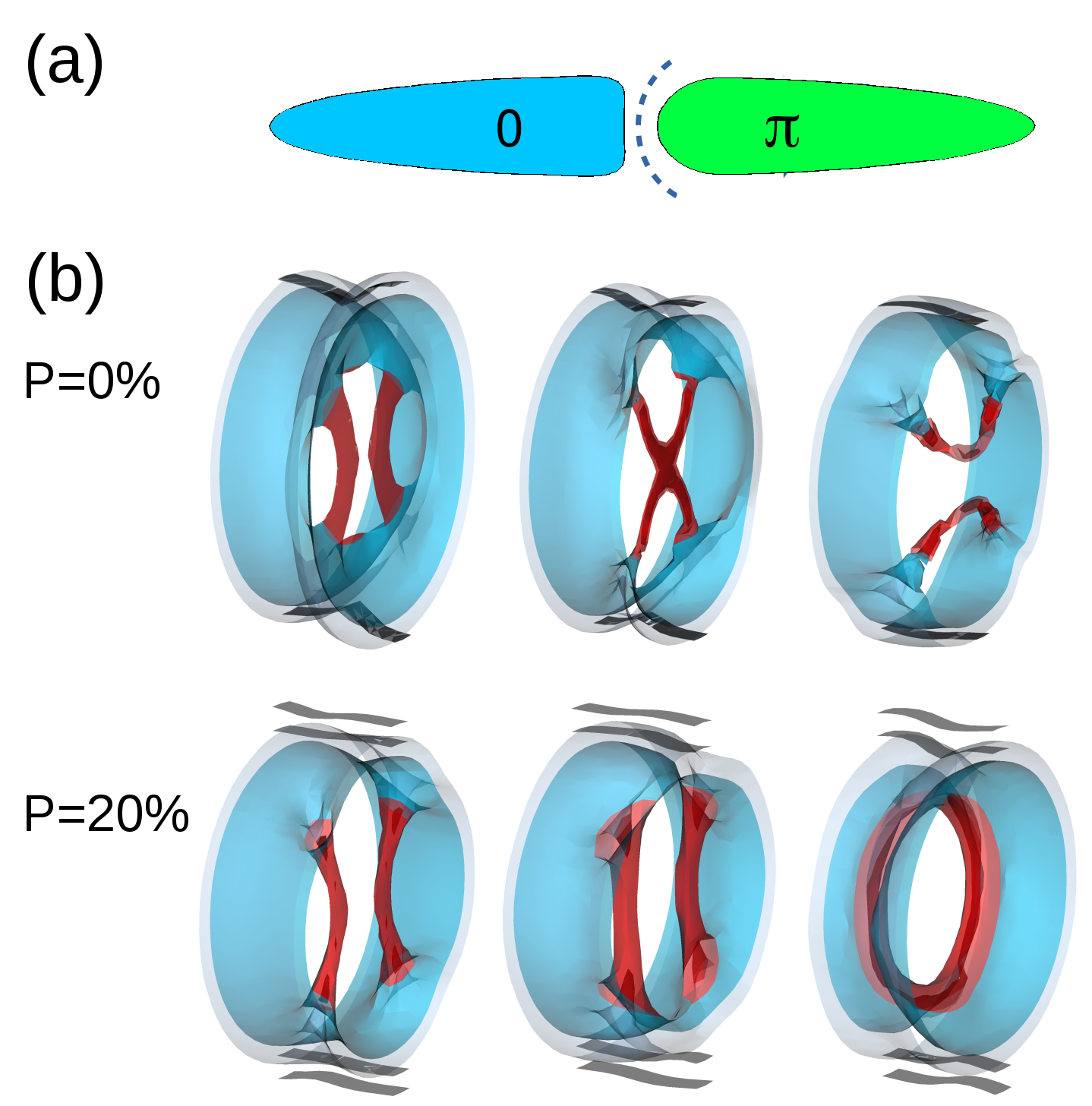}
\caption{(a) A sketch of the imprinting potential for vortex reconnection studies. (b) Consecutive stages of the time evolution of a pair of vortex lines for unpolarized $P=0\%$ (top row) and polarized $P=20\%$ (bottom row) cases. Meaning of surfaces is the same as in Fig.~\ref{fig:2}(a).
\label{fig:3}}
\end{figure}
Another process in which 
the change of topology plays an important role is the vortex reconnection, envisioned by Feynman~\cite{Feynamn}. In this process, two vortex lines approach each other, cross and exchange their fragments. The reconnection induces a deformation of the vortex lines, known as Kelvin wave, which is responsible for energy emission to the medium by the vortices. As a result the total length of two vortices after reconnection is smaller than before. The process is crucial for generation and evolution
of quantum turbulent state, defined as a chaotic motion of many quantized vortices. 
In recent years, various experiments with ultra-cold atoms of bosonic type allowed to get a deeper understanding of this phenomenon~\cite{Tsatsos}.
The results presented in this Letter indicate that the spin-imbalanced atomic gases offer a new platform for studies of the quantum turbulence phenomenon in fermionic systems, where both quantum and classical turbulence can coexist even at zero temperature limit. 

Using TDASLDA we designed the experimental setup that can be used to study the vortex reconnection process. The setup requires 
to deform the imprinting potential to a shape that resembles an arch, see Fig~\ref{fig:3}(a). The bent potential seeds immediately the snake instability that results in two parallel vortex lines with opposite circulations. In a spin symmetric system we observe the reconnection process; see Fig.~\ref{fig:3}(b). 
If we assume that topology of the order parameter plays the most important role when comes to dynamics of the defects, we can expect that similar effect occur in polarized system with $P=20\%$, because in both cases the order parameter is very similar. However, our simulation reveals a remarkable difference.
While a pair of vortex lines is created, they do not cross. The system prefers instead  
to make a step back and reform a vortex ring. Clearly, the vortex reconnection process is hindered when the vortex core is polarized. It may significantly affect the turbulent regime which requires dynamics of many tangled vortices, where reconnection processes will be inevitable. At this stage it is hard to predict to 
what extent the hindrance of reconnections will affect the turbulent dynamics in spin-imbalanced fermionic systems.
Nevertheless, we expect to observe significant changes in energy dissipation rates. In order to answer this question precisely, microscopic simulations at larger scales than those presented here are required.

\ssec{Acknowledgements} We thank A. Bulgac, J. Brand, M. Forbes and M. Zwierlein for valuable suggestions. This work was supported by the Polish National Science Center (NCN) under Contract No. UMO-2014/13/D/ST3/01940. We acknowledge PRACE for awarding us access to resource Piz Daint
based in Switzerland at Swiss National Supercomputing Centre (CSCS), decision No. 2016153479. We also acknowledge Interdisciplinary Centre for Mathematical and Computational Modelling (ICM) of Warsaw University for computing resources at Okeanos (grant No. GA67-14) and Global Scientific Information and Computing Center, Tokyo Institute of Technology for resources at TSUBAME (Project ID: hp170181). K.S. and P.M. acknowledge support from the Polish NCN grant No. UMO-2013/08/A/ST3/00708.

\ssec{Author Contributions}  The numerical codes were designed and implemented by G.W. Numerical simulations were executed by G.W. and K.S. Discussion
and interpretation of the results was carried out by G.W., K.S. and P.M. 
The codes optimization towards large scale calculations was performed by M.M. All authors contributed to writing of the manuscript.


\begin{center}
{\bf Supplementary Material for:}\\
{\bf ``Suppressed solitonic cascade in spin-imbalanced superfluid Fermi gas''}\\
\end{center}
\setcounter{figure}{3}    

\subsection{Framework}
In the simulations we employ an extension of the Density Functional Theory (DFT)~\cite{Dreizler} to time-dependent problems~\cite{Marques}. 
The Runge-Gross theorem assures that Time-Dependent DFT (TDDFT) approach is equivalent to solving many-body time-dependent Schr\"{o}dinger equation, if ``exact'' energy density functional is provided. While the exact form of the functional is typically unknown, for ultra-cold atoms at unitarity it is severely restricted by various symmetries, especially by the scale invariance property of the system. Here, we used local functional $E=\int \mathcal{H}(\bm{r})d\bm{r}$ with energy density functional, constructed by Bulgac \& Forbes, of the form ($m=\hbar=1$):
\begin{eqnarray}
    \label{eq:DF_ASLDA}
    \mathcal{H} &=&
    \alpha_{\uparrow}(n_{\uparrow},n_{\downarrow})\frac{\tau_{\uparrow}}{2} + 
    \alpha_{\downarrow}(n_{\uparrow},n_{\downarrow})\frac{\tau_{\downarrow}}{2} \nonumber \\ 
    &+&
      D(n_{\uparrow},n_{\downarrow})
    +
    g(n_{\uparrow},n_{\downarrow})\nu^{\dagger}\nu \nonumber \\ 
    &+& [1-\alpha_{\uparrow}(n_{\uparrow},n_{\downarrow})]\dfrac{\bm{j}_{\uparrow}^2}{2n_{\uparrow}}
    + [1-\alpha_{\downarrow}(n_{\uparrow},n_{\downarrow})]\dfrac{\bm{j}_{\downarrow}^2}{2n_{\downarrow}}.
\end{eqnarray}
The functional is defined through normal density $n_{\sigma}$, kinetic density $\tau_{\sigma}$, 
anomalous density $\nu$ and current density $\bm{j}_{\sigma}$, which are parametrized by functions $\{v_{n,\sigma}(\bm{r}),u_{n,\sigma}(\bm{r})\}$, called quasi-particle wave functions (qpwfs):
\begin{eqnarray}
n_{\sigma}(\bm{r}) &=& \sum_{|E_n|<E_c}\abs{v_{n,\sigma}(\bm{r})}^2 f_{\beta}(-E_n),\label{eqn:densities0}\\
\tau_{\sigma}(\bm{r}) &=& \sum_{|E_n|<E_c}\abs{\nabla v_{n,\sigma}(\bm{r})}^2 f_{\beta}(-E_n),\\
\nu(\bm{r}) &=& \sum_{|E_n|<E_c} u_{n,\uparrow}(\bm{r})v_{n,\downarrow}^{*}(\bm{r})\frac{f_{\beta}(-E_n)-f_{\beta}(E_n)}{2},\\
\bm{j}_{\sigma}(\bm{r}) &=& \sum_{|E_n|<E_c} \textrm{Im}[v_{n,\sigma}(\bm{r})\nabla v_{n,\sigma}^*(\bm{r})] f_{\beta}(-E_n),
\label{eqn:densities}
\end{eqnarray}
where subscripts $\sigma=\{\uparrow,\downarrow\}$ indicate spin, $E_{n}$ denotes quasi-particle energy and $E_c$ is energy cut-off scale. Fermi distribution function $f_{\beta}(E)=1/(\exp(\beta E)+1)$ is introduced to model temperature $T=1/\beta$ effects. Since we use only local part of the anomalous density $\nu(\bm{r})$, the cut-off is required for correct regularization of the theory. The functional together with the  regularization procedure is known in literature as Asymmetric Superfluid Local Density Approximation (ASLDA)~[18,19]. 

Terms in the functional~(\ref{eq:DF_ASLDA}) have very
simple physical meaning. The first two represents the kinetic energy contribution of particles with spins $\uparrow$ and $\downarrow$. The effective mass $\alpha_{\sigma}$ of the particle depends on local polarization $p(\bm{r})=\frac{\nup(\bm{r})-\ndown(\bm{r})}{\nup(\bm{r})+\ndown(\bm{r})}$ and guarantees that correct limit is attained for $\nup\gg\ndown$, where the problem reduces to the polaron problem~\cite{PolaronProblem}. The next term represents the normal interaction energy $D$, while the term proportional to $|\nu|^2$ is due to Cooper pairing. Last two terms restore Galilean invariance of the functional. Functions $\alpha_{\sigma}$, $D$ and $g$
were adjusted to available Quantum Monte Carlo results for the unitary Fermi gas (UFG), both polarized and unpolarized, as well as being in superfluid and normal state. The functional was validated against experimental results including both static and dynamic properties. Explicit forms of functions $\alpha_{\sigma}$, $D$ and $g$ are given in~[18].

It is important to stress that the functional~(\ref{eq:DF_ASLDA})  
offers a description of superfluididity beyond Bogoliubov de-Gennes (BdG) approximation.
Indeed, the description can be identical to BdG if we set $\alpha_{\uparrow}=\alpha_{\downarrow}=1$, $D=0$ and $g=\textrm{const}$. It means that in BdG all interaction effects are modelled through pairing correlations and the normal state is described as free Fermi gas. This is very crude approximation in the case of the UFG, which is known to be strongly interacting system, even above the critical temperature. In ASLDA, interaction effects in the normal state are always present due to $D(n_{\uparrow},n_{\downarrow})$ term.

The real-time dynamics are given by equations, which have generic form similar to the Time-Dependent Bogolubov-de Gennes (TDBdG) equations:
\begin{equation}\label{eq:tddft}
  i  \frac{\partial}{\partial t}
  \begin{pmatrix}
    u_{n,\uparrow}(\bm{r},t) \\ 
    v_{n,\downarrow}(\bm{r},t)
  \end{pmatrix} = 
  \begin{pmatrix}
    h_{\uparrow}(\bm{r},t) & \Delta(\bm{r},t) \\
    \Delta^*(\bm{r},t)& -h_{\downarrow}^*(\bm{r},t) 
  \end{pmatrix}
  \begin{pmatrix}
    u_{n,\uparrow}(\bm{r},t) \\ 
    v_{n,\downarrow}(\bm{r},t)
  \end{pmatrix}.
\end{equation}
Here we refer to them as Time-Dependent ASLDA (TDASLDA). The single particle hamiltonian $h_{\sigma}=k_{\sigma}+U_{\sigma}+V_{\textrm{ext}}$ consists of the kinetic part $k_{\sigma}$ (containing spatial derivatives), the mean field part $U_{\sigma}$ and the external potential $V_{\textrm{ext}}$. 
The mean field potentials $U_{\sigma}(\bm{r},t)$ and pairing potential $\Delta(\bm{r},t)$ are derived from the energy density functional as a functional derivative $U_{\sigma}=\frac{\delta E}{\delta n_{\sigma}}$ and $\Delta=-\frac{\delta E}{\delta \nu^*}$. They explicitly depend on local densities $n_{\sigma}(\bm{r})$, $\tau_{\sigma}(\bm{r})$, etc., that are again defined by qpwfs through the formulas~(\ref{eqn:densities0})--(\ref{eqn:densities}), where missing components of qpwfs can be found via symmetry relation: $u_{n,\uparrow}\rightarrow v_{n,\uparrow}^*$, $v_{n,\downarrow}\rightarrow u_{n,\downarrow}^*$ and $E_n\rightarrow -E_n$. 

\subsection{Numerical implementation}
The set of TDASLDA equations~(\ref{eq:tddft}) represents a system of coupled, complex, nonlinear, time-dependent, 3D partial differential equations (PDEs), which has to be solved self-consistently. The main problem arises from
the fact that in the presence of superfluidity ($\Delta\neq 0$) one has to evolve a huge number of quasi-particle (2-component) wave-functions, which in practical simulations is of the order
of $10^5 - 10^6$.  All of these equations are coupled through densities. The qpwfs are represented on conceptually simple,
discrete, 3D spatial lattice with lattice constant $a$, and $N_i$ lattice points fixed in each spatial
direction ($i=x,y,z$), with periodic boundary conditions.
The TDASLDA is extremely well suited for spatial lattice formulation since the single-particle potentials ($U_{\sigma}$ and $\Delta$) are local in space. The coupling
between neighboring lattice points comes only from the Laplacian or gradient operations. Spatial derivatives are calculated using FFT, which is one of the most accurate method for functions with periodic boundary conditions~\cite{SpectralMethods}. 

For time integration we used Adams-Bashforth-Moulton (predictor-corrector) scheme of 5th order. Note that all qpwfs fluctuate in time with a typical oscillating factor $\exp(-i E_n t)$, where $E_n$ is a quasi-particle energy. Since we evolve all states with quasi-particle energies from  
the interval $[-E_c,E_c]$, the evolved states exhibit both very slow as well as very rapid oscillations in time.
In order to increase accuracy of the time integration, in each integration step we subtract from the quasi-particle solution the typical frequency of oscillations, which in practice modifies 
the equation to the form:  
$i \dot{\varphi_n} = (\hat{H}-\avg{H}_n)\varphi_n$, where $\varphi_n=(u_{n,\uparrow}, v_{n,\downarrow})^{T}$ and $\hat{H}$ is quasi-particle Hamiltonian, and $\avg{H}_n$ is ``instantaneous'' quasi-particle energy. Although this step introduces significant numeric cost to the calculations, it greatly improves accuracy of the time integration. 

The integration time step $\Delta t$ is taken to be $\Delta t\,E_{\textrm{max}}=0.05$, where $E_{\textrm{max}}=\pi^2/2m$ is the estimation of the maximum energy that can be resolved on 
spatial lattice with the lattice spacing $a=1$. We checked that further decreasing of the integration time step does not  
affect the results.
During the computation process we were monitoring the energy and the particle number conservation (separately for spin-up and spin-down particles). They were constant (within small uncertainty) for all generated trajectories.  

The code uses multiple Graphical Porcessing Units (GPUs) and is based on mixed CUDA~\cite{cuda} and MPI (Message Passing Interface)~\cite{mpi} programming model. Qpwfs are distributed between GPUs and each GPU stores information only about part of qpwfs. All operations related to qpwf are executed by GPUs. MPI communication is needed only for computation of densities (\ref{eqn:densities0})--(\ref{eqn:densities}). The code has been implemented from scratch. The cuFFT library is used to perform fast Fourier transforms needed for computations of derivatives. The code has been optimized for performance (the shortest time of one time step) when running on computers equipped with at least hundreds of GPUs. The most time consuming part is related to the FFT execution, which takes about 50\% of the running time. 

\subsection{Simulation details}
Numerical experiments were executed on Piz Daint machine---presently the fastest European supercomputer~\cite{Top500}. This machine allowed us to simulate dynamics of about $600$ strongly interacting atoms in three dimension (3D) without any symmetry restrictions. 
Simulations were executed on spatial lattice of size $32\times 32\times 128$ with the lattice constant $a=1$. Number of evolved states was about $125,000$.
The gas of atoms was trapped in an external harmonic potential of the form:
\begin{equation}
 V_{\textrm{ext}}(x,y,z)=\dfrac{m\omega_x^2 x^2}{2}+\dfrac{m\omega_y^2 y^2}{2}+\dfrac{m\omega_z^2 z^2}{2}+Cy^3,
\end{equation} 
where $\frac{\omega_x-\omega_y}{\omega_x+\omega_y}=0.015$, and $\omega_x\cong 4\omega_z$. The harmonic potential is supplemented by anharmonicity term $Cy^3$ which simulates influence of the gravity~[13] and it brakes reflection symmetry of the problem with respect to $y$-axis. Anharmonicity parameter was set $C=0.002\,C_0$, where $C_0=\frac{m\omega_y^2}{2R}$, $R=\sqrt{\frac{2\mu}{m\omega_y^2}}$ is Thomas-Fermi radius and $\mu=0.37\varepsilon_F$. The Fermi energy $\varepsilon_F=\frac{\hbar^2(6\pi^2n_{\uparrow})^{2/3}}{2m}$ is set by density of spin-up component (majority component) in the center of the cloud at the beginning of simulation. We worked with very cold gas, with temperature $T/\varepsilon_F=0.01$, well below superfluid-to-normal phase transition. In simulations we used $\varepsilon_F\approx 0.5$ and $E_c\approx 10\varepsilon_F$.

\begin{figure}[h]
\includegraphics[width=\columnwidth]{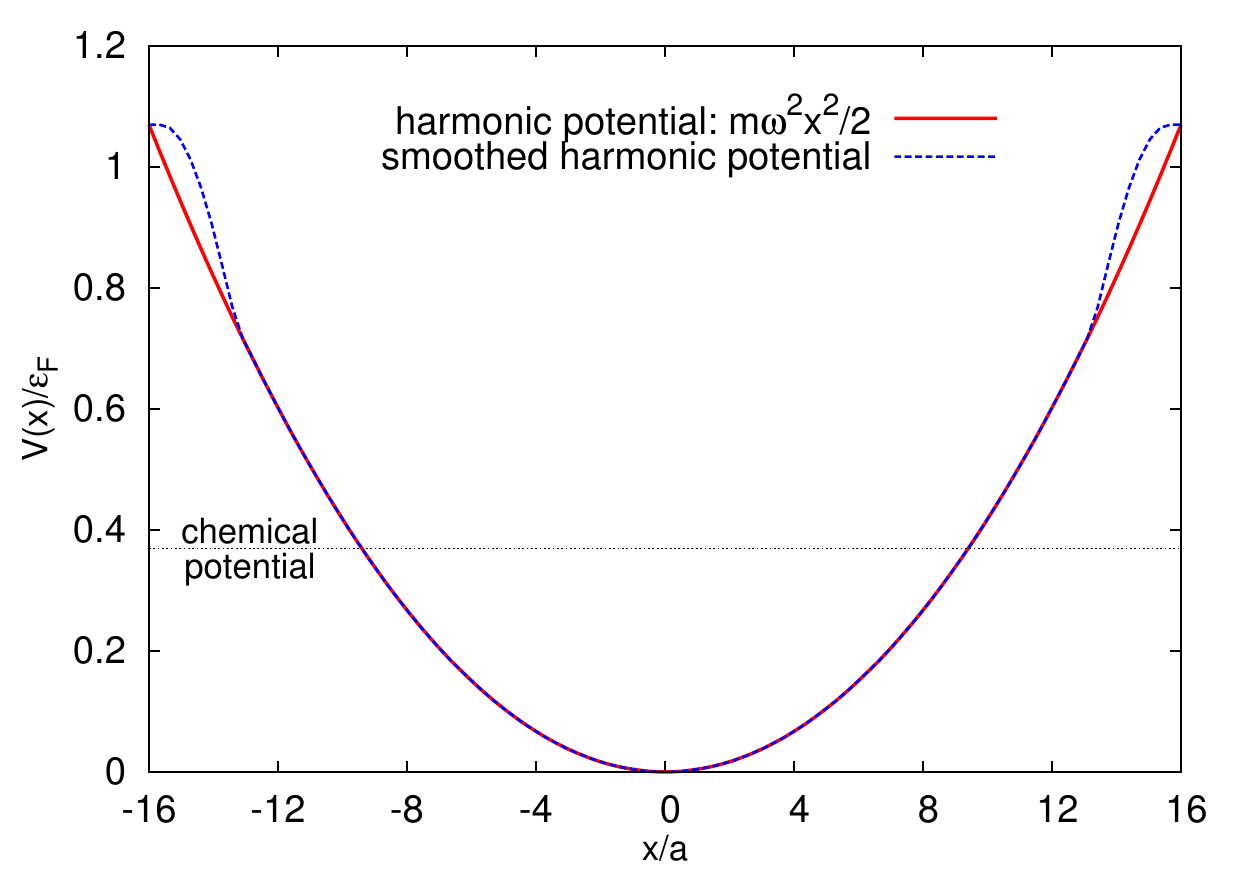}
\caption{The harmonic potential along $x$ axis for box of size $N_x=32$ (red solid line). This harmonic potential has a discontinuous first derivative at the edges of the box. In order to avoid numerical errors, in calculations we used modified version which changes the potential only near the edges (blue dashed line). Horizontal dashed line demonstrates location of the chemical potential of the majority component.\label{fig:ho}}
\end{figure}

In the simulations the periodic boundary conditions were imposed in each direction. The harmonic potential $m\omega^2 x^2/2$ does not fulfill periodicity requirement, which may generate significant errors in computation process. To avoid these errors, in practical realization we modified the potential near edges of the box in such a way that its derivatives are smooth and continuous over the entire domain; see Fig~\ref{fig:ho}. The value of the potential at the boundary was set to be about $3$ times larger than chemical potential of majority component. 
Ratios of the chemical potentials to the trap angular frequency are: $\bar{\mu}/\hbar \omega_{x} \approx 3.5$, $3.2$, $3.0$, $2.9$ for $P=0\%$, $20\%$, $40\%$ and $50\%$, respectively, where $\bar{\mu} = (\mu_{\uparrow}+\mu_{\downarrow})/2$.

\begin{figure}[h]
\includegraphics[width=\columnwidth]{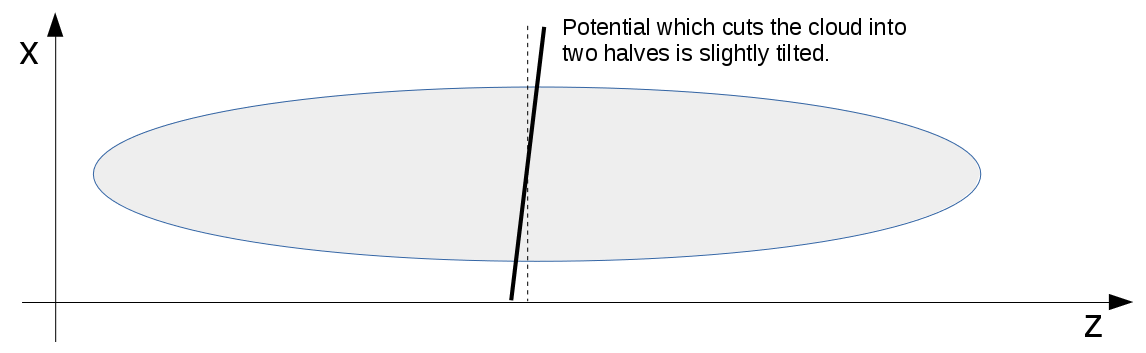}
\caption{ Demonstration of tilting of the knife potential.\label{fig:tilt}}
\end{figure}

The phase imprint procedure consists of adding constant potential to one side of the cloud for time interval $T_{\rm phase}=\pi/2U_0$. We have used the potential form of
$V_{\rm impr.}=U_0\Theta(x,z)$, where the smoothed step function $\Theta(x,z)$
(uniform in $y$ direction) is given by
\begin{eqnarray}
&&\Theta(x,z) = \nonumber\\[2mm]
&&\hspace{-3mm}\left\lbrace 
\begin{array}{ll}
   0,                      & z \leqslant z_0(x)-1,                  \\
   s(z-z_0(x)+1,2),        & z_0(x)-1 < z < z_0(x)+1,               \\
   1,                      & z_0(x)+1 \leqslant z \leqslant 0.9L_z, \\
   1-s(z-0.9L_z,0.1L_z),   & z \geqslant 0.9L_z,
\end{array}
\right.\nonumber\\[-3mm]
\label{eq:V_impr}
\end{eqnarray}
where $L_z=aN_z$ is the length of the numerical box in $z$ direction.
$s(z,w)$ is a function which smoothly varies from $0$ to $1$ in an interval $[0,w]$:
\begin{equation}
 s(z,w)=\dfrac{1}{2}+
 \dfrac{1}{2}\tanh\left[\tan\left(  \frac{\pi z}{w}-\frac{\pi}{2} \right) \right].
 \label{eq:switch}
\end{equation}
The potential $V_{\rm impr.}$ advances the phase of the order parameter by $\pi$ for $z>z_0(x)$ over the time duration $T_{\rm phase}$. The phase imprinting potential changes rapidly in the vicinity of $z=z_0=0$, which 
means that particles experience a strong kick in that region and it may cause numerical instability of the algorithm. In order to avoid this problem, the phase imprint procedure is accompanied with a repulsive knife-edge potential:
\begin{equation}\label{eq:knife_edge}
V_{\textrm{knife}} = U_1\exp \left[ -\frac{(z-z_0(x))^2}{2\sigma^2} \right],
\end{equation}
which separates the cloud into two halves. The barrier height is the same as the value of the harmonic potential at the box boundary, while the width was set to $\sigma=2$. The knife generates vacuum in the vicinity of $z=z_0$, and undesired effects related to  
the sharp change of the imprint potential can be avoided. In spin-imbalanced system, it allows for spin excess component to reside in the region where the dark soliton is going to be created. Otherwise, the slow spin diffusion process will not allow for fast equilibration of the dark soliton state, after phase imprinting.
To account for experimental imperfections, we assume that the phase-imprinting beam is slightly tilted which is reflected by the presence of function $z_0(x)=a_0 (x-L_x/2)$, where we chose $a_0=1/16$; see also Fig~\ref{fig:tilt}. Here, we choose the tilt in $x-z$ plane, and it also serves as a mechanism for the reflection symmetry breaking with respect to $x$-axis. As noted in~\cite{Scherpelz}, the tilt is a reliable mechanism for setting the vortex line orientation in a spin-symmetric system. 
Reflection symmetry with respect to $z$-axis is broken due to the phase imprint procedure, and therefore we work with the system where all spatial symmetries are broken. 

For the study of vortex reconnections, we proposed to bend (instead of the tilt) the Gaussian knife potential [cf. Fig.~3(a) of the main text]. It was achieved by modifying the knife position $z_0(x)$ as follows:
\begin{equation}
z_0(x) = k \frac{4}{L_x^2}\Bigl( x - \frac{L_x}{2} \Bigr)^2 + \frac{L_z}{2},
\end{equation}
where $k=5$ is used in our simulations. As is easily seen, the knife position
along $x$ direction is a quadratic function about $x=\frac{L_x}{2}$, reaching
largest displacement at the edge of the box, $z_0(0)=z_0(L_x)=k+L_z/2$.

\subsection{Initial state preparation}

The simulation needs to 
start from a nontrivial initial configuration. The initial configuration is obtained as a solution of the static version of Eq.~(\ref{eq:tddft}), where we replace $i \frac{\partial}{\partial t}\rightarrow E_n$. Instead of solving ASLDA equations in 3D, which is prohibitively expensive, we followed a different strategy presented in~\cite{QuantumFriction}. Our initial state preparation procedure
consists of two distinct steps: 1) generation of a static solution
in a 2D system confined in $V_{\rm ext}(x,y)=m\omega_x^2x^2/2+
m\omega_y^2y^2/2+Cy^3$; 2) dynamic implementation of the harmonic
trap in $z$ direction $V_{\rm ext}(z)=m\omega_z^2z^2/2$, cutting
the cloud into two halves by the knife-edge potential (\ref{eq:knife_edge})
and the phase imprinting for one of the clouds.

\begin{figure*}[t]
\includegraphics[width=\textwidth, trim=0 290 0 50, clip]{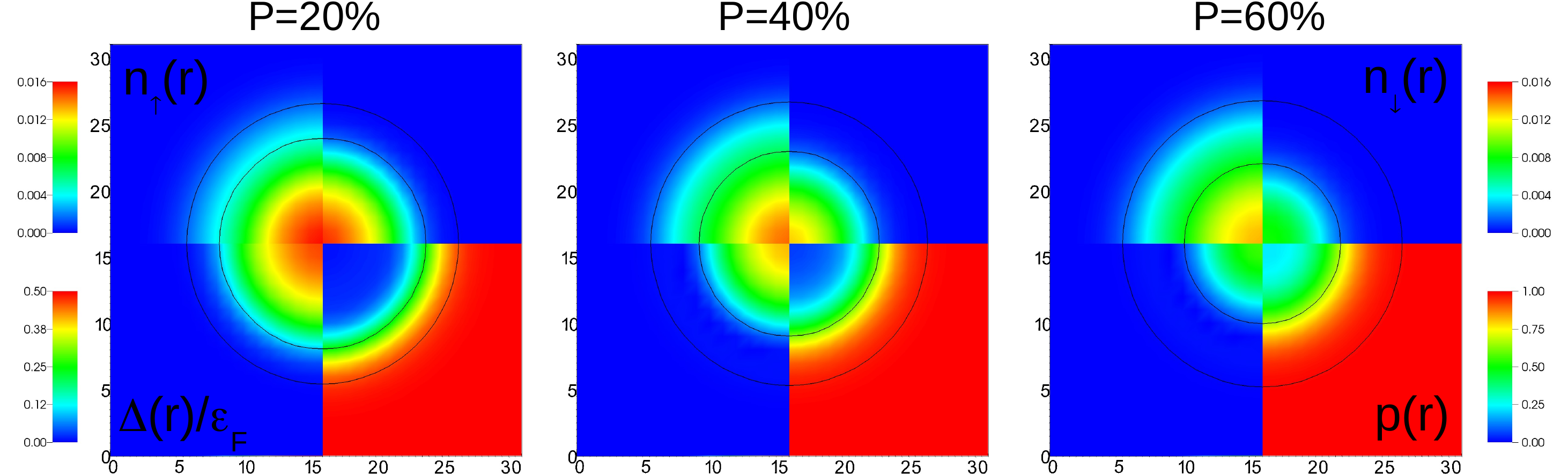}
\caption{
Initial configurations for polarizations $P=20\%$, $40\%$ and $60\%$. In each quarter of the panel different quantities are presented: density distributions of majority  $n_{\uparrow}$ (upper left) and minority $n_{\downarrow}$ (upper right) components, paring field $\Delta$ (lower left) and local polarization $p$ (lower right). Outer (inner) black circle shows region where density of majority (minority) component decreases to value $0.0016$. Between these two circles the normal phase resides, while the superfluid phase is located inside the cloud. 
\label{fig:initstates}}
\end{figure*}

First, we generated a solution of a 2D system that is uniform in
$z$ direction. Namely, the qpwfs are assumed to have the following
structure, \textit{e.g.} $u_{n,\sigma} (\boldsymbol{r})=u_{n,\sigma}(x,y)e^{ik_zz}$.
This assumption allows us to replace the 3D problem with $N_z$th
2D problems for each $k_z$, \textit{i.e.}, 0, $\pm2\pi/L_z$, $\dots$, $+\pi/a$.
In Fig.~\ref{fig:initstates} we show the 2D solutions for three cases: $P=20\%$, $40\%$ and $60\%$. It is clearly visible that the initial configurations exhibit the phase separation pattern, \textit{i.e.} a fully-paired superfluid component ($p(\bm{r})\approx 0$) is located inside the cloud, while unpaired particles being in a normal state ($\Delta(\bm{r})\approx 0$) create the outer shell. 

Second, starting from the 2D solution, we introduce the external
potentials to generate appropriate initial states for our numerical
experiments. The details of the timing setups are:
\begin{itemize}
\setlength{\leftskip}{24mm}
\setlength{\parskip}{2mm}
\setlength{\itemsep}{0mm}
\item [\textrm{[0,     150]:}]{turning on $V_{\rm ext}(z)$,}
\item [\textrm{[200,   350]:}]{turning on $V_{\rm knife}$,}
\item [\textrm{[400,   410]:}]{turning on $V_{\rm impr.}$,}
\item [\textrm{[500,   510]:}]{turning off $V_{\rm knife}$ and $V_{\rm impr.}$,}
\item [\textrm{[510,  1000]:}]{observation of the dynamics,}
\end{itemize}
where those numbers in the squared parentheses indicate a multiple of time
and the Fermi energy, $t\,\varepsilon_{\rm F}$. All the dynamic changes of
the external potentials were introduced smoothly using the switching function
(\ref{eq:switch}). During the time interval $[0, 400]$,
the time evolution was supplemented with the quantum friction potential
that suppresses irrotational currents, reducing the total energy of
the system during the real time evolution. As a result we obtained low energy states which can be treated as a good approximation of the ground state of the 3D problem~\cite{QuantumFriction}. 

\begin{figure}[t]
\includegraphics[width=\columnwidth]{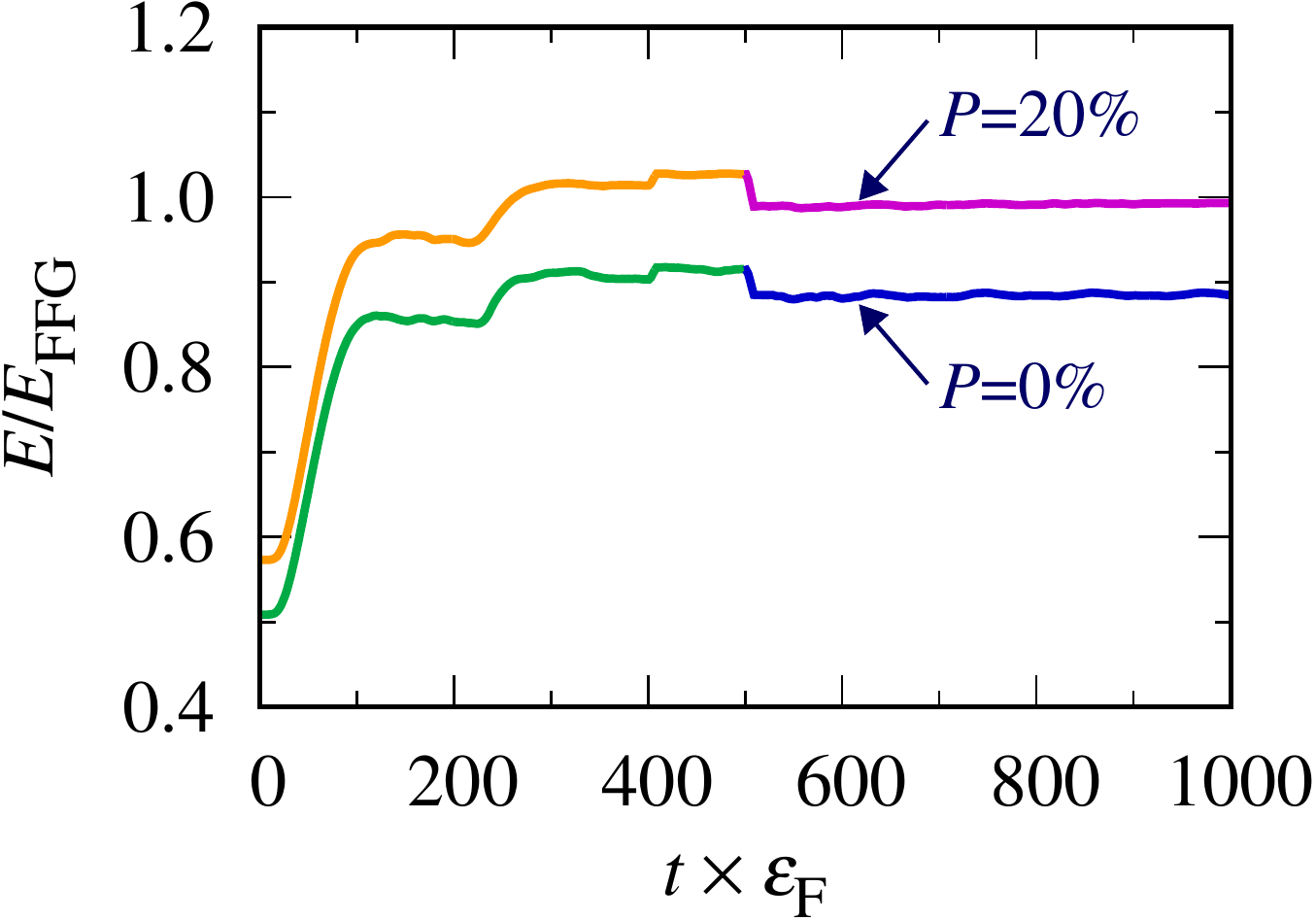}
\caption{
Time evolution of the total energy divided by the energy of the
free Fermi gas (FFG), $E/E_{\rm FFG}$, in two cases, $P=0\%$ and 20\%.
$E_{\rm FFG}$ is computed with respect to the local number densities
$n_\sigma(\boldsymbol{r})$ at the center of the cloud at $t=0$, \textit{i.e.},
before introducing $V_{\rm ext}(z)$.
\label{fig:energy}
}
\end{figure}

In Fig.~\ref{fig:energy}, we show the total energy relative to the free
Fermi gas value as a function of $t\,\varepsilon_{\rm F}$, for two
representative cases, $P=0\%$ and 20\%. From the figure, one finds
three rapid increases of the total energy at times around $t\,\varepsilon_{\rm F}=
[0, 150]$, $[200,350]$ and $[400,410]$. They correspond to the introduction
of the harmonic trap in $z$ direction, the Gaussian knife potential and
the phase imprinting step function, respectively. The decreasing trend
in interval $t\,\varepsilon_{\rm F}=[150,200]$ and $[350,400]$ is because
of the cooling by the quantum friction. After the removal of $V_{\rm knife}$
and $V_{\rm impr.}$ ($t\,\varepsilon_{\rm F}>510$), we see that the total
energy is nicely conserved, as it should be. This demonstrates the accuracy
of our TDASLDA simulations, during the simulation period as long as
$t\,\varepsilon_{\rm F}=$1,000.

\subsection{Visualization of results}
We provide movies showing dynamics of the system and offering a better insight into physics involved
in presented phenomena. 
Below we provide sample frames from different types of visualizations with explanation of the meaning of depicted elements. The visualization starts just before removing the knife potential (\textit{i.e.} the initial state preparation is not shown).

\subsection{Visualization 1}
\begin{figure*}[th]
\includegraphics[width=\textwidth, trim=50 145 50 50, clip]{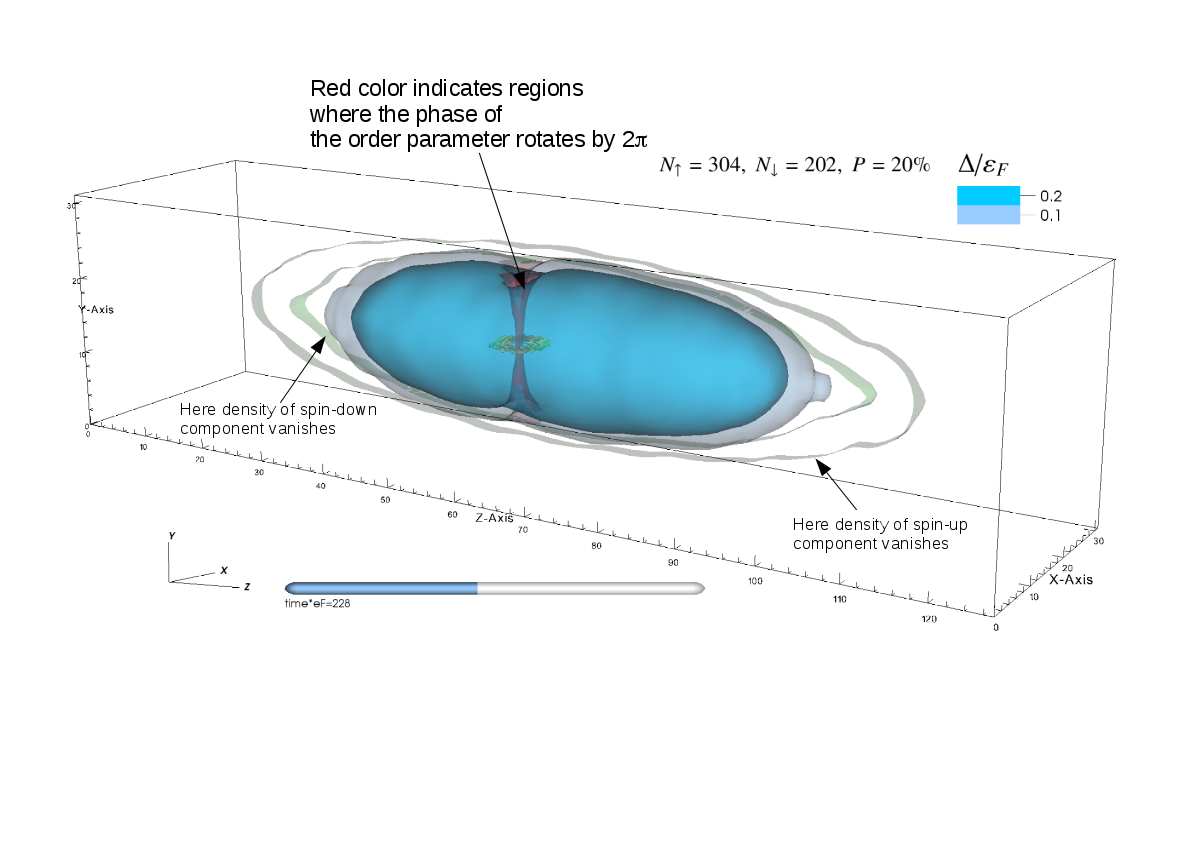}
\caption{Sample frame together with meaning of plotted items for type 1 visualization. 
\label{Sfig:1}}
\end{figure*}
This visualization shows the evolution of topology of the order parameter $\Delta(\bm{r})$; see Fig.~\ref{Sfig:1} for a sample frame. Two 3D contours are plotted by blue color for $\Delta(\bm{r})/\varepsilon_F=0.1$ and $0.2$. Vortex detection algorithm has been applied in order to highlight by red color regions where: i) absolute value of the order parameter goes to zero, ii) phase of the order parameter rotates by $2\pi$ as we go along closed contour (in vicinity of considered region). Appearance of these two criteria is regarded as a signature of the presence of a quantized vortex. For clarity, we also show velocity field of majority component $\bm{v}_{\uparrow}(\bm{r})$ in the vicinity of the defect. Note that there are rare instances where the detection criteria fail with the proper vortex detection. This occurs
usually for regions close to the condensate edge, where the order parameter naturally vanishes and the phase has very complicated structure. The visualization also shows boundaries of spin-up and spin-down clouds. These are defined as the points where the density of the component decreases approximately $10$ times in comparison to the central value. They are plotted by two gray layers surrounding the condensate. Between these shells the highly polarized gas in normal state resides. Spin-up fermions are always the majority component and therefore are forming the outer layer. 
In the case of spin-balanced system the two layers overlap. 

\subsection{Visualization 2}
\begin{figure*}[h]
\includegraphics[width=\textwidth, trim=50 100 50 60, clip]{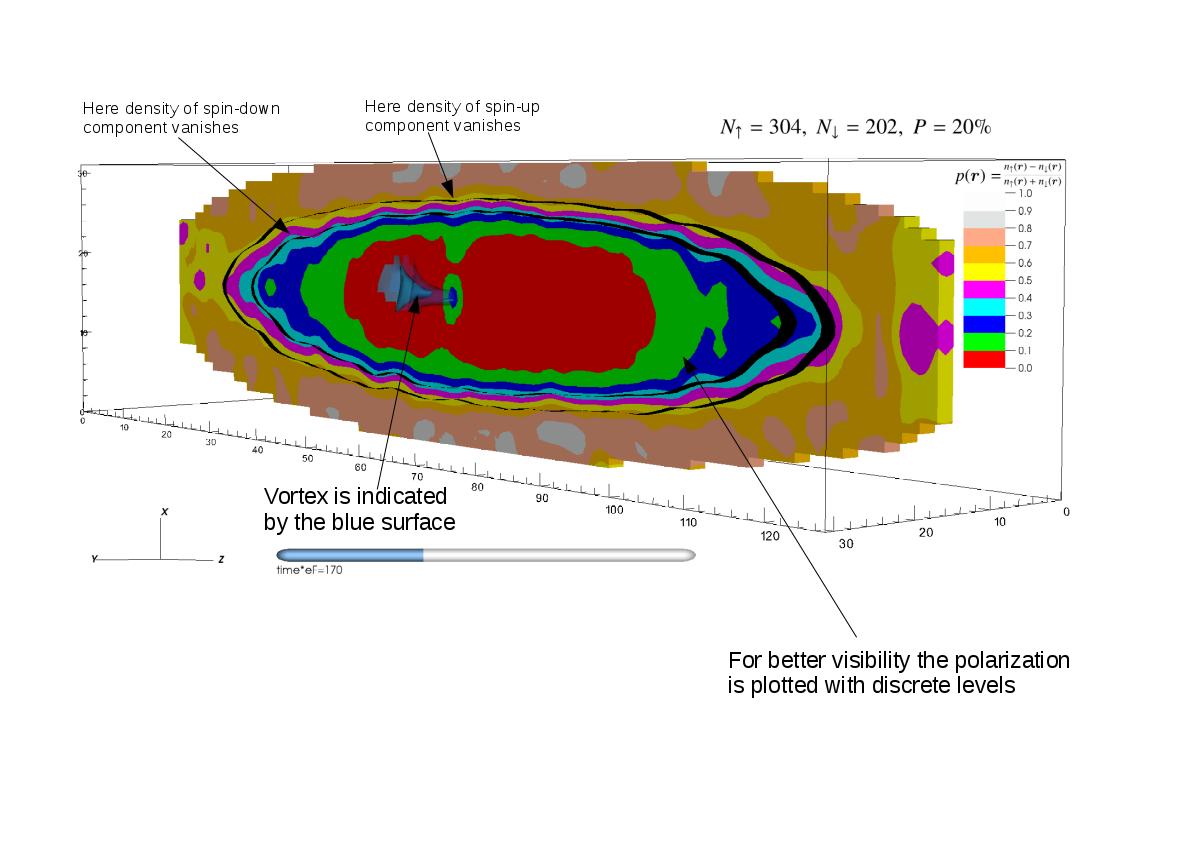}
\caption{Sample frame together with meaning of plotted items for type 2 visualization.
\label{Sfig:2}}
\end{figure*}

This visualization provides insight into time evolution of the local polarization of the gas $p(\bm{r})=\frac{n_{\uparrow}(\bm{r})-n_{\downarrow}(\bm{r})}{n_{\uparrow}(\bm{r})+n_{\downarrow}(\bm{r})}$; see Fig.~\ref{Sfig:2} for a sample frame. This type of visualization is provided only for spin-polarized runs. The polarization is plotted only for section in $x-z$ plane through center of the box. For better visibility the polarization is plotted by discrete levels. Similar to Visualization 1, we provide also information about spin-up and spin-down cloud sizes, plotted by black lines. If a quantized vortex is detected, it is shown by blue surface.

\subsection{Visualization 3}
\begin{figure*}[h]
\includegraphics[width=\textwidth, trim=45 25 50 60, clip]{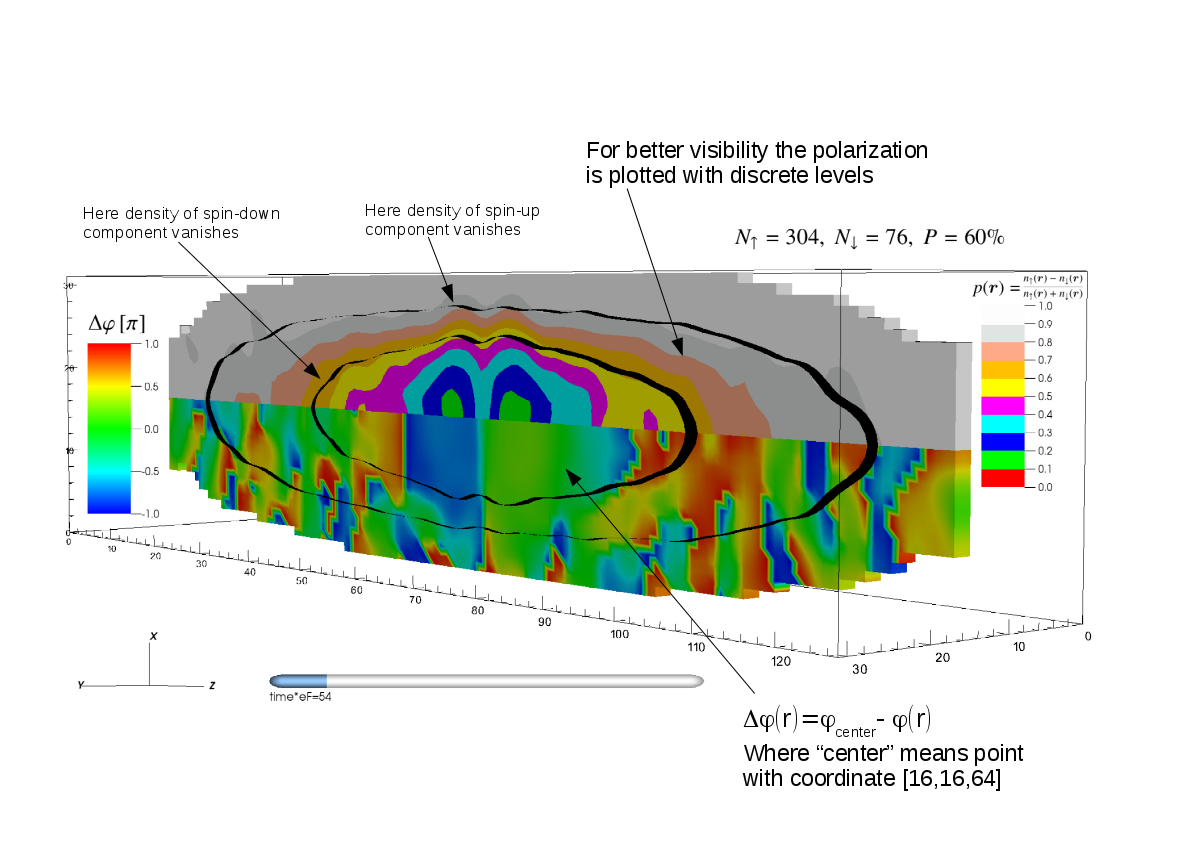}
\caption{Sample frame together with meaning of plotted items for type 3 visualization.
\label{Sfig:3}}
\end{figure*}

This visualization is very similar to the Visualization 2, but now the additional information about the phase of the order parameter $\Delta(\bm{r})$ is provided in the bottom part of the section; see Fig.~\ref{Sfig:3} for a sample frame. The visualization is very useful for demonstrating stability of the dark soliton for runs with $P=50\%$ and $P=60\%$. The soliton is characterized by a sharp jump of the phase by $\pi$ in nodal plane, while for a quantized vortex the phase rotates by $2\pi$ as we move along the contour encapsulating the vortex core. For better visualization we plot the phase difference $\Delta\varphi$ defined as difference between phase in selected point and reference point which is located in the center of the box. Then, the soliton appears as plane where phase changes sharply from $0$ (green color) to $-\pi$ (blue color) or $\pi$ (red color). 
Note that the sharp change of phase from $-\pi$ to $\pi$ (rapid change of blue into red color) do not mean discontinuity of the phase, as the phase is periodic function with $2\pi$ period.

\subsection{List of movies}

\noindent Below we provide list of movies. All movies are also accessible on YouTube. 
\begin{description}
\item[Movie 1:] Run for a spin-balanced system, $P=0\%$. Number of atoms is $N_{\uparrow}=N_{\downarrow}=304$.\\[2mm]
{\it Visualization 1}: File=\verb|Movie1-P0.0-Vis1.mp4|, YouTube=\url{https://youtu.be/OCUDHW2GAGc}\\[1mm]
{\it Visualization 4}: File=\verb|Movie1-P0.0-Vis4.mp4|, YouTube=\url{https://youtu.be/BAew87ry84c}\\[1mm]
{\it Visualization 5}: File=\verb|Movie1-P0.0-Vis5.mp4|, YouTube=\url{https://youtu.be/CemhQrWDUa4}\\
These two additional visualizations (4 and 5) show evolution  of the order parameter in vicinity of the nodal plane created by the phase imprint procedure. Frames from these visualizations were used for Fig.~1 in the main text.

\item[Movie 2:] The same as Movie 1, except that the knife potential is not titled in $x-z$ plane. The movie demonstrates that the tilt of the knife edge changes only orientation of the vortex line in final state, while it does not changes qualitatively the results of Movie~1.\\[2mm]
{\it Visualization 1}: File=\verb|Movie2-P0.0-Vis1.mp4|, YouTube=\url{https://youtu.be/ZCK9I6rxiF0}

\item[Movie 3:] Run for a spin-imbalanced system, $P=20\%$. Number of atoms is $N_{\uparrow}=304$ and $N_{\downarrow}=202$. The full solitonic cascade is observed. Frames from the movie were used to create Fig.~2(a) of the main text.\\[2mm]
{\it Visualization 1}: File=\verb|Movie3-P0.2-Vis1.mp4|, YouTube=\url{https://youtu.be/FCKLvm0dJlU}\\[1mm]
{\it Visualization 2}: File=\verb|Movie3-P0.2-Vis2.mp4|, YouTube=\url{https://youtu.be/PjSsUYRKY40}

\item[Movie 4:] Run for a spin-imbalanced system, $P=40\%$. Number of atoms is $N_{\uparrow}=304$ and $N_{\downarrow}=130$. The cascade ends with the vortex ring creation. The ring rapidly dissipates its energy
and is ejected from the cloud.\\[2mm]
{\it Visualization 1}: File=\verb|Movie4-P0.4-Vis1.mp4|, YouTube=\url{https://youtu.be/U4WXb8-VHG4}\\[1mm]
{\it Visualization 2}: File=\verb|Movie4-P0.4-Vis2.mp4|, YouTube=\url{https://youtu.be/WV6dqjLcx24}\\[1mm]
{\it Visualization 3}: File=\verb|Movie4-P0.4-Vis3.mp4|, YouTube=\url{https://youtu.be/XBNdyMqqOgg}

\item[Movie 5:] Run for a spin-imbalanced system, $P=50\%$. Number of atoms is $N_{\uparrow}=304$ and $N_{\downarrow}=100$. The dark soliton is stabilized and do not decay into neither a vortex ring nor a vortex line. It propages towards left edge of the cloud and is ejected from the system.\\[2mm]
{\it Visualization 1}: File=\verb|Movie5-P0.5-Vis1.mp4|, YouTube=\url{https://youtu.be/jIQJG_95Q2Y}\\[1mm]
{\it Visualization 2}: File=\verb|Movie5-P0.5-Vis2.mp4|, YouTube=\url{https://youtu.be/oJprsWdJPgM}\\[1mm]
{\it Visualization 3}: File=\verb|Movie5-P0.5-Vis3.mp4|, YouTube=\url{https://youtu.be/losbJzYP5HI}

\item[Movie 6:] Run for a spin-imbalanced system, $P=60\%$. Number of atoms is $N_{\uparrow}=304$ and $N_{\downarrow}=76$. Result are qualitatively similar to result of Movie 5.\\[2mm]
{\it Visualization 1}: File=\verb|Movie6-P0.6-Vis1.mp4|, YouTube=\url{https://youtu.be/WEtSHhobFTM}\\[1mm]
{\it Visualization 2}: File=\verb|Movie6-P0.6-Vis2.mp4|, YouTube=\url{https://youtu.be/v5_w8ZQhB7M}\\[1mm]
{\it Visualization 3}: File=\verb|Movie6-P0.6-Vis3.mp4|, YouTube=\url{https://youtu.be/JBJ1ix3klHc}

\item[Movie 7:] Run for a spin-balanced system, $P=0\%$. Number of atoms is $N_{\uparrow}=N_{\downarrow}=76$. The condensate has similar size as in Movie 6. The movie demonstrates that the condensate is still large enough to allow a dark soliton to decay into a vortex line.\\[2mm]
{\it Visualization 1}: File=\verb|Movie7-P0.0-Vis1.mp4|, YouTube=\url{https://youtu.be/bgfUuGoj-5I}

\item[Movie 8:] Run for a spin-balanced system, $P=0\%$. Number of atoms is $N_{\uparrow}=N_{\downarrow}=304$. The knife potential is bend according to formula (11) from Methods section. The movie demonstrates vortex reconnection process.\\[2mm]
{\it Visualization 1}: File=\verb|Movie8-P0.0-Vis1.mp4|, YouTube=\url{https://youtu.be/rFAFWXH9LBU}

\item[Movie 9:] Run for a spin-imbalanced system, $P=20\%$. Number of atoms is $N_{\uparrow}=304$ and $N_{\downarrow}=202$. The knife potential is bend as in Movie 8. The movie demonstrates that in the polarized medium the vortex reconnection process is hindered.\\[2mm]
{\it Visualization 1}: File=\verb|Movie9-P0.2-Vis1.mp4|, YouTube=\url{https://youtu.be/doH_NT0PNHU}\\[1mm]
{\it Visualization 2}: File=\verb|Movie9-P0.2-Vis2.mp4|, YouTube=\url{https://youtu.be/mXne654R0GQ}

\end{description}
\end{document}